\documentclass[%
 amsmath,amssymb,
 aps,
 prb,
]{revtex4-2}

\usepackage{graphicx}
\usepackage{dcolumn}
\usepackage{bm}
\usepackage{float}
\usepackage{algorithm}
\usepackage{algpseudocode}
\usepackage{amsmath,amssymb}

\begin{document}

\preprint{APS/123-QED}

\title{Efficient recursive vertex-contraction algorithm for the evaluation of Matsubara sums in finite-temperature many-body perturbation theory}

\author{Amel KAIBI}
 \email{amel.kaibi@univ-tebessa.dz}
 \author{Mohamed Amine TAG}
 \email{mohamed.tag@univ-tebessa.dz}
 \affiliation{Laboratory of Computational Physics and Quantum Phenomena, Cheikh Larbi Tebessi university, Tebessa, Algeria}

\date{\today}

\begin{abstract}
We present the Contraction Algorithm for Matsubara Sums (CAMS), a recursive vertex-contraction method that evaluates the internal Matsubara frequency sums of connected Feynman or Hugenholtz diagrams. The method is not restricted to two-body interactions: it applies to $p$-body vertices, to vacuum diagrams, and to self-energy diagrams with an external frequency, and the internal frequencies may be either fermionic or bosonic. The result is a closed expression in the propagator energies and the inverse temperature. The recursion runs over subsets of connected vertex that contain a fixed root vertex, so it does not enumerate the $n_v!$ time orderings of the vertices. For a diagram with $n_v$ vertices and $n$ edges, the computational cost is $O(n\,2^{\,n_v-1})$, where $n=2n_v$ for a two-body system.

\end{abstract}

\maketitle

\section{\label{sec:Into}Introduction}

Finite-temperature many-body perturbation theory provides a fundamental framework for investigating interacting quantum systems in thermal equilibrium \cite{matsubara1955,abrikosov1975,fetter2003,mahan2000,negele1998}. Within the imaginary-time formalism, thermodynamic quantities and correlation functions are expressed as perturbative expansions involving Feynman diagrams \cite{feynman1949,luttinger1960}, with each diagram contributing a multidimensional sum over internal Matsubara frequencies and, generally, an integration over internal momenta. Diagrams of the Goldstone and Hugenholtz type organize the same expansions for fermionic many-body systems \cite{goldstone1957,hugenholtz1957,bloch1958}, and the finite-temperature theory continues to be developed and tested in electronic-structure applications \cite{kohn1960,hirata2013,santra2017,hirata2021}. Although the diagrammatic formulation offers a systematic organization of perturbative contributions, the analytical evaluation of the associated frequency sums becomes increasingly challenging as the perturbation order increases.

For a given Feynman diagram, the internal Matsubara frequencies are not independent. Their values are constrained by energy conservation at the interaction vertices, and the resulting frequency summations are coupled through the propagators connecting these vertices. Consequently, the direct evaluation of a high-order diagram may require several nested summations, whose analytical treatment becomes cumbersome even for relatively simple interaction models. Developing systematic procedures to eliminate these frequency variables is therefore an important problem in computational many-body physics. The problem is made more pressing by diagrammatic Monte Carlo methods, which sample high-order diagrammatic series stochastically \cite{prokofev1998,prokofev2007,vanhoucke2010,kozik2010,vanhoucke2012} and have been extended to determinant-based sampling of the series \cite{rossi2017,simkovic2019}. In these schemes, each frequency sum that can be performed analytically reduces the dimension of the space that must be sampled.

Several approaches have been developed to address this difficulty. Conventional analytical techniques rely primarily on contour integration, residue calculus, and partial-fraction decompositions to evaluate Matsubara sums \cite{abrikosov1975,fetter2003,mahan2000}. These methods are well established, but their direct application to high-order diagrams can involve increasingly complicated algebraic expressions. Other approaches exploit the structure of the thermal averages and of the diagrams to organize the frequency summations. In particular, Gaudin's formulation of the statistical Wick theorem \cite{gaudin1960} underlies the standard diagrammatic rules at finite temperature, and graphical methods built on it connect Matsubara summations with tree structures of Feynman diagrams\cite{belghit2026automated,tag2024finite}. Thermal operator representations have also been developed to relate finite-temperature diagrammatic expressions to their zero-temperature counterparts \cite{brandt2005,brandt2006}.

More recently, algorithmic approaches have been introduced to automate the analytical evaluation of Matsubara frequency sums. Among these, Algorithmic Matsubara Integration (AMI), developed by Taheridehkordi, Curnoe, and LeBlanc, employs repeated applications of the residue theorem to construct analytical expressions for diagrammatic frequency contributions \cite{taheridehkordi2019}. This approach has enabled the systematic treatment of high-order diagrammatic expansions and their applications to interacting fermionic systems, including the Hubbard model. Subsequent developments have combined it with the optimal grouping of diagrams \cite{taheridehkordi2020optimal}, extended its use to real-frequency calculations of response functions \cite{taheridehkordi2020susc,vucicevic2020}, and provided a dedicated software library \cite{elazab2022}. These developments demonstrate the importance of algorithmic treatments of Matsubara sums in modern computational many-body theory.

In this work, we introduce an alternative approach based on recursive contractions between the vertices of vacuum Feynman diagrams. Rather than treating frequency summations as a sequence of independent algebraic operations, the proposed method exploits the connectivity of the diagram and the frequency-conservation structure associated with its vertices. Starting from a selected vertex, referred to as the root, the algorithm progressively contracts neighboring vertices until the entire diagram has been reduced to a terminal state. Each contraction generates a finite collection of contributions characterized by hyperbolic functions of specific combinations of propagator energies. These contributions are combined through a recursive relation involving an energy denominator associated with the current contraction state.

A central feature of the method is that the contraction process can be formulated entirely in terms of the topology of the diagram and the incidence relations of its propagators. To achieve this, the vertices and propagators are encoded using binary masks, allowing the relevant graph operations, including the identification of neighboring vertices, the updating of contracted subsets, and the construction of energy combinations, to be performed through bitwise operations. The resulting algorithm separates the construction of the contraction hierarchy from its numerical evaluation. The hierarchy is generated in a forward traversal from the root vertex, while the corresponding contributions are evaluated recursively in the reverse direction, starting from the fully contracted diagram.

An additional computational advantage arises from the representation of intermediate configurations by their contracted vertex subsets. Different contraction sequences can lead to the same subset of vertices. By storing each distinct configuration only once, the algorithm avoids the repeated evaluation of equivalent intermediate states. This feature transforms the recursive procedure into a subset-based dynamic programming scheme and provides a natural framework for investigating its computational complexity as the diagram order increases.

The proposed method is applicable to diagrams represented through their vertex-propagator incidence structure, including the Feynman or Hugenholtz representations \cite{feynman1949,hugenholtz1957}. Once the Matsubara frequency summations have been eliminated, the resulting expressions depend only on the propagator energies and the inverse temperature. The remaining momentum integrations can therefore be treated independently using analytical or numerical integration techniques, including Monte Carlo sampling \cite{vanhoucke2010,kozik2010,rossi2017}.

The objective of this paper is to establish the mathematical formulation of the recursive vertex-contraction procedure and to develop an efficient computational implementation for evaluating finite-temperature Matsubara sums. Particular attention is given to the construction of the contraction hierarchy, the algebraic structure of the resulting numerator and denominator factors, and the reuse of intermediate contraction states. The accuracy and computational performance of the method are then investigated through benchmark calculations and comparisons with independent evaluations of Matsubara sums.

\section{\label{sec:Matsubara}Matsubara sums}

The contribution of each Feynman or Hugenholtz diagram with $n$ internal propagators is
\begin{equation}
    I(\{\nu_{\rm ext}\})=\frac{1}{\beta^{L}}\sum_{\{\nu_{\rm int}\}}\prod_{i=1}^{n}G(i\nu_i),
    \label{integral}
\end{equation}
where $\{\nu_{\rm int}\}$ are the independent internal Matsubara frequencies, $\{\nu_{\rm ext}\}$ is the set of external frequencies, and
\begin{equation}
    L=n-n_v+1
\end{equation}
is the number of independent internal frequencies of a connected diagram. For a two-body vacuum diagram, $L=n_v+1$. The bare Green's function is
\begin{equation}
    G(i\nu_k) = \frac{1}{i\nu_k - \varepsilon_k}.
    \label{green}
\end{equation}
The internal frequencies are either fermionic, $\nu=(2m+1)\pi/\beta$, or bosonic, $\nu=2m\pi/\beta$, where $m$ is an integer.

\section{Basic principle of the vertex-contraction procedure}

\begin{figure}[ht]
        \centering
        \includegraphics[width=0.88\columnwidth]{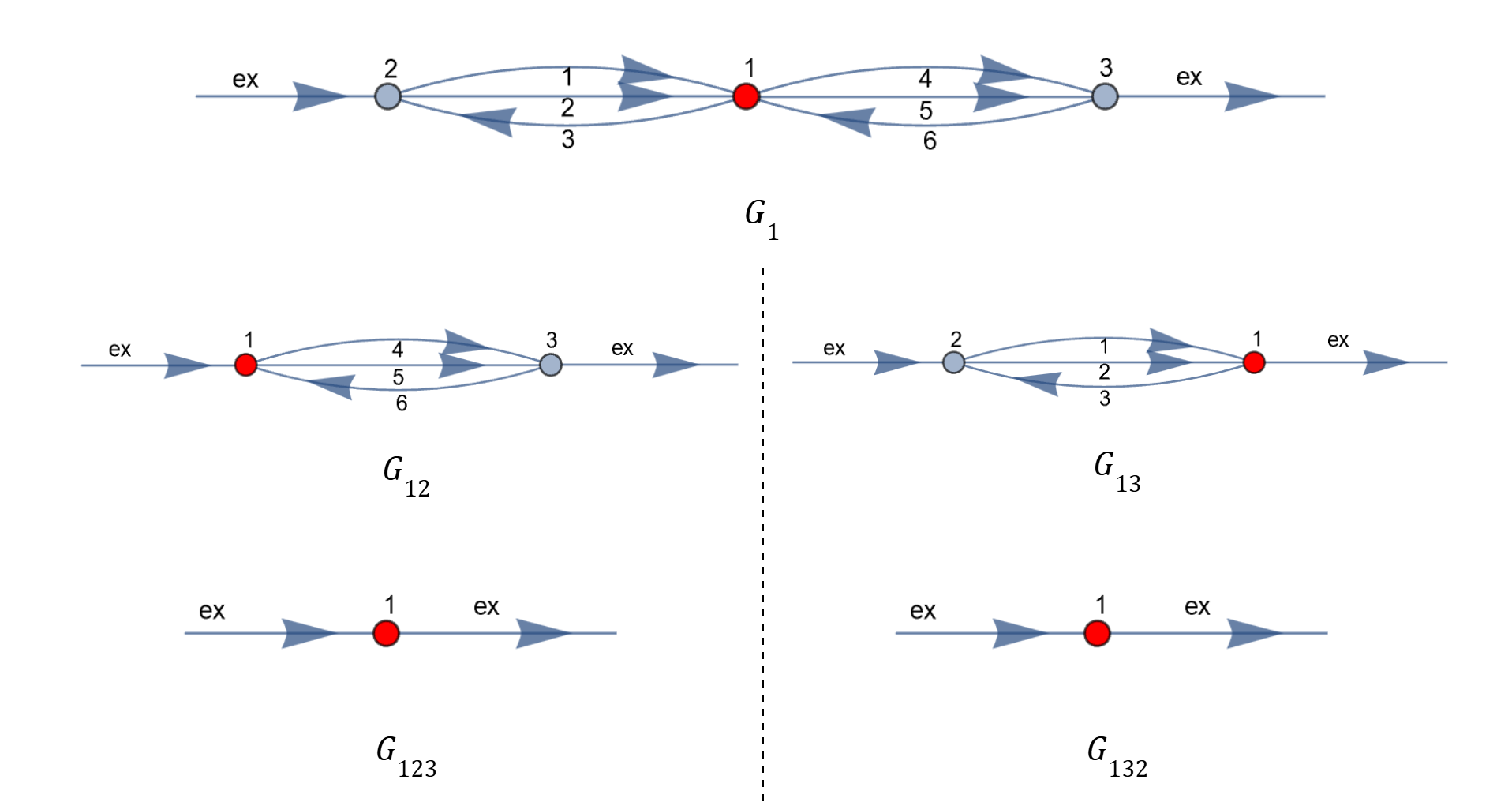}
        \caption{Contraction process for four external frequencies.}
        \label{fig1}
\end{figure}

Consider the diagram illustrated in Fig.~\ref{fig1}, consisting of three interaction vertices connected by four internal propagators. The propagators are labeled 1,2,3,4,5, and 6, with corresponding energies $\varepsilon_{p_1}$, $\varepsilon_{p_2}$, $\varepsilon_{p_3}$, $\varepsilon_{p_4}$, $\varepsilon_{p_5}$  and $\varepsilon_{p_6}$. In addition, the diagram contains one external legs carrying the external frequency $z_{ext} = i v_{ext}$.

The central vertex, labeled 1, is selected as the root of the contraction procedure. The basic idea of the method is to progressively eliminate the vertices of the diagram by contracting the root vertex with one of its neighboring vertices. Each contraction eliminates the propagators connecting the two selected vertices and produces a reduced diagram with updated external legs and energy labels.

In the present example, the root vertex 1 has two neighboring vertices, 2 and 3. Consequently, two possible contractions can be performed.

The first possibility consists of contracting vertex 1 with vertex 2, as illustrated in Fig.~\ref{fig1}. This operation eliminates the three internal propagators labeled 1,2 and 3, leaving a reduced diagram containing vertex 3, the internal propagators 4, 5 and 6, and the external frequency $z_{ext}$.

Alternatively, vertex 1 can be contracted with vertex 3. In this case, the propagators labeled 4, 5 and 6 are eliminated, and the resulting reduced diagram contains vertex 1, the propagators 1,2 and 3, and the external frequency $z_{ext}$.

The contribution of the original diagram is obtained by summing the contributions associated with these two possible contractions. Each contribution consists of a thermal contraction factor, determined by the propagators connecting the contracted vertices, multiplied by the contribution of the corresponding reduced diagram. Both terms share an energy denominator determined by the incoming and outgoing propagators incident on the original root vertex.

We first consider the contraction of vertices 1 and 2. The three connecting propagators are labeled 1, 2 and 3. According to their orientations, propagators 1 and 2 are incoming to vertex 2, whereas propagator 3 is outgoing. The corresponding thermal contraction factor is therefore
\begin{equation}
K_{12}=
\frac{
\cosh\left[\dfrac{\beta}{2}
\left(\varepsilon_{p_3}-\varepsilon_{p_1}-\varepsilon_{p_2}\right)\right]
}{
4\cosh\left(\dfrac{\beta\varepsilon_{p_1}}{2}\right)
\cosh\left(\dfrac{\beta\varepsilon_{p_2}}{2}\right)\cosh\left(\dfrac{\beta\varepsilon_{p_3}}{2}\right)
}.
\end{equation}
Similarly, for the contraction of vertices 1 and 3, the connecting propagators are 4, 5 and 6. Propagators 4, 5 are outgoing from vertex 2, whereas propagator 6 is incoming. The corresponding contraction factor reads
\begin{equation}
K_{13}=
-\frac{
\cosh\left[\dfrac{\beta}{2}
\left(\varepsilon_{p_4}+\varepsilon_{p_5}-\varepsilon_{p_6}\right)\right]
}{
4\cosh\left(\dfrac{\beta\varepsilon_{p_4}}{2}\right)\cosh\left(\dfrac{\beta\varepsilon_{p_5}}{2}\right)
\cosh\left(\dfrac{\beta\varepsilon_{p_6}}{2}\right)
}.
\end{equation}
The denominator associated with the root vertex 1 is determined by all four internal propagators incident on this vertex. Taking into account their orientations, we obtain
\begin{equation}
\Delta E_1=
\varepsilon_{p_3}+\varepsilon_{p_4}+\varepsilon_{p_5}
-\varepsilon_{p_1}-\varepsilon_{p_2}-\varepsilon_{p_6}.
\end{equation}
It is important to distinguish this denominator from the energy combinations appearing in the thermal contraction factors. Whereas $K_{12}$ and $K_{13}$ depend only on the propagators connecting the corresponding pairs of vertices, $\Delta E_1$ involves all incoming and outgoing internal propagators of the root vertex.

Combining the two possible contractions, the contribution of the original diagram can be expressed as
\begin{equation}
    G_0=
\frac{
K_{12}G(p_4,p_5,p_6,z_{ext})
+
K_{13}G(p_1,p_2,p_3, z_{ext})
}{
\Delta E_1
}.
\label{4}
\end{equation}
Here, the functions $G(p_4,p_5,p_6,z_{ext})$ and $G(p_1,p_2,p_3,z{ext})$ represent the contributions associated with the two reduced diagrams obtained after the respective contractions.

The contraction procedure is applied recursively to each reduced diagram until all interaction vertices have been eliminated.

Consider first the reduced diagram obtained by contracting vertices 1 and 2. The remaining internal propagators are 4, 5 and 6, while the external frequency is $z_{ext}$. The remaining vertex can now be contracted with the external legs, resulting in the energy denominator
\begin{equation}
\Delta E_{23}=
\varepsilon_{p_4}+\varepsilon_{p_5}
-\varepsilon_{p_6}-z_{ext}.
\end{equation}
The corresponding thermal contraction factor is $K_{23}$. Consequently, the contribution of this reduced diagram is
\begin{equation}
G(p_4,p_5,p_6,z_{ext})=
-\frac{
\cosh\left[\dfrac{\beta}{2}
(\varepsilon_{p_4}+\varepsilon_{p_5}-\varepsilon_{p_6})\right]
}{
4\cosh\left(\dfrac{\beta\varepsilon_{p_4}}{2}\right)\cosh\left(\dfrac{\beta\varepsilon_{p_5}}{2}\right)
\cosh\left(\dfrac{\beta\varepsilon_{p_6}}{2}\right)
\left(\Delta E_{23}\right)
}.
\label{6}
\end{equation}
Similarly, for the alternative reduced diagram obtained by contracting vertices 1 and 3, the remaining propagators are 1, 2 and 3, and the external frequency is $z_{ext}$. The corresponding energy denominator is
\begin{equation}
\Delta E_{32}=
\varepsilon_{p_3}+z_{ext}
-\varepsilon_{p_1}-\varepsilon_{p_2}.
\end{equation}
Its contribution is therefore
\begin{equation}
    G(p_1,p_2,p_3,z_{ext})=
\frac{
\cosh\left[\dfrac{\beta}{2}
(\varepsilon_{p_3}-\varepsilon_{p_1}-\varepsilon_{p_2})\right]
}{
4\cosh\left(\dfrac{\beta\varepsilon_{p_1}}{2}\right)\cosh\left(\dfrac{\beta\varepsilon_{p_2}}{2}\right)
\cosh\left(\dfrac{\beta\varepsilon_{p_3}}{2}\right)
\left(\Delta E_{32}\right)
}.
\label{8}
\end{equation}
Substitution of Eqs.~(\ref{6}) and (\ref{8}) into Eq.~(\ref{4}) gives the complete analytical contribution of the original three-vertex diagram.

This example illustrates the recursive nature of the proposed method. At each stage, a selected vertex is contracted with one of its neighboring vertices, generating a reduced diagram and an associated algebraic contribution. The external frequencies are retained in the reduced diagram and enter the energy denominators of subsequent contractions.

\section{General formulation of the vertex-contraction rule}

We now formulate the contraction procedure for a general vacuum Feynman diagram.

Let \(v_r\) denote the selected root vertex and \(v_j\) one of its neighboring vertices. We define the set of propagators connecting these two vertices by \(\mathcal E(v_r,v_j)\). This set is divided into outgoing and incoming propagators, according to their orientations relative to the root vertex:
\[
\mathcal E(v_r,v_j)
=
\mathcal E_{\mathrm{out}}(v_r,v_j)
\cup
\mathcal E_{\mathrm{in}}(v_r,v_j).
\]
We denote the total number of connecting propagators by \(m=|\mathcal E(v_r,v_j)|\), and the number of incoming propagators by \(n_{\mathrm{in}}=|\mathcal E_{\mathrm{in}}(v_r,v_j)|\). The energy combination associated with the contracted propagators is
\begin{equation}
\Delta E_{rj}^{\mathrm{conn}}
=
\sum_{p\in\mathcal E_{\mathrm{out}}(v_r,v_j)}
\varepsilon_p
-
\sum_{p\in\mathcal E_{\mathrm{in}}(v_r,v_j)}
\varepsilon_p.
\label{9}
\end{equation}
The thermal factor generated by the contraction is, for fermion propagators,
\begin{equation}
    K(v_r,v_j)=
\frac{(-1)^{n_{\mathrm{in}}}}
{\displaystyle
\prod_{p\in\mathcal E(v_r,v_j)}
\cosh\left(\frac{\beta\varepsilon_p}{2}\right)}
\begin{cases}
\displaystyle
\cosh\left(
\frac{\beta\Delta E_{rj}^{\mathrm{conn}}}{2}
\right),
& m\ \text{odd},\\[10pt]
\displaystyle
\sinh\left(
\frac{\beta\Delta E_{rj}^{\mathrm{conn}}}{2}
\right),
& m\ \text{even},
\end{cases}
\label{10}
\end{equation}
and, for boson propagators,
\begin{equation}
    K(v_r,v_j)=
\frac{(-1)^{m}}
{\prod_{p\in\mathcal E(v_r,v_j)}
\sinh\left(\frac{\beta\varepsilon_p}{2}\right)}\sinh\left(
\frac{\beta\Delta E_{rj}^{\mathrm{conn}}}{2}
\right).
\label{eq11}
\end{equation}
The overall numerical normalization associated with the contractions is specified separately in the complete diagrammatic expression (Sec.~\ref{sec:cams}).

The energy denominator is constructed from all propagators incident on the selected root vertex in the current reduced diagram. Denoting the outgoing and incoming incident propagator sets by \(\mathcal E_{\mathrm{out}}(v_r)\) and \(\mathcal E_{\mathrm{in}}(v_r)\), respectively, we define
\begin{equation}
    \Delta E_r=
\sum_{p\in\mathcal E_{\mathrm{out}}(v_r)}
\varepsilon_p
-
\sum_{p\in\mathcal E_{\mathrm{in}}(v_r)}
\varepsilon_p.
\label{11}
\end{equation}
For reduced diagrams, the external frequency labels must also be included in the corresponding incoming or outgoing energy sums, according to their orientations. Thus, the denominator is updated after every contraction to account for the external frequencies inherited from the eliminated vertices.

Let \(\mathcal N(v_r)\) denote the set of neighboring vertices of the root in the current reduced diagram. The contribution associated with the contraction of the root with all its possible neighbors can then be written as
\begin{equation}
    G(S)=
\frac{1}{\Delta E_r(S)}
\sum_{v_j\in\mathcal N(v_r)}
K(v_r,v_j)\,
G(S\cup\{v_j\}).
\label{12}
\end{equation}
Here, \(S\) denotes the set of vertices already contracted, and \(G(S\cup\{v_j\})\) represents the contribution of the reduced diagram obtained by incorporating the neighboring vertex \(v_j\) into the contracted set. The recursion terminates when all interaction vertices have been contracted, for which the terminal contribution is defined as unity.

Equation~(\ref{12}) constitutes the basic recursive relation underlying the proposed algorithm. It expresses the contribution of a given contraction state as a sum over all admissible next contractions, with each term determined by a local thermal factor and the contribution of the corresponding reduced diagram.

\section{Computational representation and recursive evaluation}

The recursive contraction procedure can be implemented efficiently by representing the diagram through its vertex-propagator incidence structure.

For a diagram containing \(n_v\) interaction vertices, each vertex is assigned a unique binary label. A contracted subset of vertices is then represented by an integer bitmask. This representation allows the identification of already contracted vertices, the determination of neighboring vertices, and the construction of updated contraction states using bitwise operations.

The propagator incidence information is encoded in binary masks as well. These masks are used to determine the incoming and outgoing propagator sets associated with each contraction state, from which the corresponding thermal factors and energy denominators are constructed.

The computational procedure is divided into two stages. First, the contraction hierarchy is generated in a forward traversal, starting from the selected root vertex. At each level, all admissible neighboring vertices are identified and the corresponding reduced states are constructed. Second, the algebraic contributions are evaluated in the reverse direction, beginning with the terminal state and proceeding toward the initial root state.

An important feature of this implementation is that different contraction sequences may lead to the same contracted vertex subset. Such equivalent intermediate states need not be evaluated repeatedly. By storing each distinct subset only once, the algorithm avoids redundant computations and provides a subset-based dynamic programming implementation of the recursive contraction procedure.

This graph-based formulation separates the topological construction of the contraction hierarchy from the numerical evaluation of the resulting expressions. Once the hierarchy has been generated, the Matsubara frequency summations can be evaluated for arbitrary values of the propagator energies and inverse temperature. The remaining momentum integrations can subsequently be performed independently using analytical or numerical techniques.

\section{\label{sec:cams}Contraction Algorithm for Matsubara Sums (CAMS)}

\begin{algorithm}[H]
\caption{Analytic forward generation of denominators and numerators}
\label{alg:forward}
\begin{algorithmic}[1]
\Require Graph with vertices $\{1,\dots,n_v\}$ and edges $\{e_1,\dots,e_n\}$; root vertex $r$
\Ensure Level lists $\mathrm{Den}_k,\ \mathrm{Num}_k,\ \mathrm{Succ}_k$ for $k=2,\dots,n_v$
\State \textbf{Precompute} for every vertex $i$:
\Statex \hspace{\algorithmicindent} $d_i \gets$ mask of all incident edges
\Statex \hspace{\algorithmicindent} $p_i \gets$ mask of all outgoing edges
\Statex \hspace{\algorithmicindent} $N_i \gets$ mask of the vertices adjacent to $i$
\State $\mathrm{Level}_1 \gets \{\, b_r \mapsto (N_r,\ d_r,\ p_r) \,\}$
\For{$k = 2$ \textbf{to} $n_v$}
    \State $\mathrm{Level}_k \gets \emptyset$
    \ForAll{$S \in \mathrm{Level}_{k-1}$}
        \State $(\nu,\delta,\pi) \gets \mathrm{Level}_{k-1}[S]$ \Comment{$\nu$: neighbours, $\delta$: cut edges, $\pi$: outgoing edges}
        \State $P \gets \pi \wedge \delta$;\quad $Q \gets \delta \oplus P$
        \State $\mathrm{Den}_k[S] \gets (P,Q)$ \Comment{cut edges leaving / entering $S$}
        \ForAll{vertices $v$ in $\nu$}
            \State $m \gets \delta \wedge d_v$ \Comment{edges joining $v$ to $S$}
            \State $P_v \gets \pi \wedge m$;\quad $Q_v \gets m \oplus P_v$
            \State append $(P_v,Q_v)$ to $\mathrm{Num}_k[S]$
            \State $S' \gets S \vee b_v$;\quad append $S'$ to $\mathrm{Succ}_k[S]$
            \If{$k<n_v$ \textbf{and} $S' \notin \mathrm{Level}_k$}
                \State $\mathrm{Level}_k[S'] \gets \big((\nu \vee N_v)\setminus S',\ \ \delta \oplus d_v,\ \ \pi \vee p_v\big)$
            \EndIf
        \EndFor
    \EndFor
\EndFor
\State \Return $\mathrm{Den}_k,\ \mathrm{Num}_k,\ \mathrm{Succ}_k$ for $k=2,\dots,n_v$
\end{algorithmic}
\end{algorithm}

The CAMS procedure is organized in two stages, given in Algorithms~\ref{alg:forward} and~\ref{alg:backward}. The first stage, the analytic forward generation (Algorithm~\ref{alg:forward}), depends only on the topology of the diagram, which is encoded by the incidence masks of its edges, and never uses the propagator energies. Starting from the subset that contains only the root vertex, it extends each subset by one adjacent vertex at a time, so that after $k-1$ steps the subsets contain $k$ vertices and the whole diagram is eventually contracted into a single vertex set. Every subset is stored as a bit mask, and the propagators crossing the boundary of the subset are obtained from the masks by a single exclusive-or with the edge mask of the added vertex, since an edge joining two vertices that are both inside the subset appears twice and cancels. These cut edges are split into those leaving and those entering the subset, which gives, for each subset, the pair of edge lists that defines its denominator, and, for each possible added vertex, the pair of edge lists that defines the numerator of that transition together with the label of the resulting subset. The output of this stage is a set of integer lists that can be stored and reused.

The second stage, the numerical backward evaluation (Algorithm~\ref{alg:backward}), takes these lists together with a set of energies $\varepsilon_1,\dots,\varepsilon_{n}$ and the inverse temperature $\beta$. It assigns the value $1$ to the full vertex set and then moves backward through the levels, from subsets with $n_v-1$ vertices down to the root. The value of each subset is the sum, over its transitions, of the numerator factor multiplied by the value of the successor subset, divided by the denominator of the subset. For fermionic lines the numerator factor is a hyperbolic cosine or sine of $\beta/2$ times the signed energy sum, depending on the parity of the number of cut edges, with a sign given by the number of entering edges; for bosonic lines it is always a hyperbolic sine, with sign $(-1)^{|P|+|Q|}$. The Matsubara sum is finally obtained from the value of the root subset by multiplying by $2^{-L}$, with $L=n-n_v+1$ the number of independent internal frequencies ($L=n_v+1$ for a two-body diagram), and by the product $\prod_e \operatorname{sech}(\beta\varepsilon_e/2)$ for fermions or $\prod_e 1/\sinh(\beta\varepsilon_e/2)$ for bosons.

\begin{algorithm}[H]
\caption{Numerical backward evaluation}
\label{alg:backward}
\begin{algorithmic}[1]
\Require Output of Algorithm~\ref{alg:forward}; energies $\varepsilon_1,\dots,\varepsilon_{n}$; inverse temperature $\beta$; fermi=true
\Ensure Value $\mathcal{I}$ of the Matsubara sum
\Function{$\Sigma$}{$P,Q$}
    \State \Return $\sum_{e\in P}\varepsilon_e-\sum_{e\in Q}\varepsilon_e$
\EndFunction
\Function{$h$}{$P,Q$}
    \State $x \gets \beta\,\Sigma(P,Q)/2$
    \If{fermi}
        \If{$|P|+|Q|$ is odd}
            \State \Return $(-1)^{|Q|}\cosh x$
        \Else
            \State \Return $(-1)^{|Q|}\sinh x$
        \EndIf
    \Else
        \State \Return $(-1)^{|P|+|Q|}\sinh x$
    \EndIf
\EndFunction
\State $G[\{1,\dots,n_v\}] \gets 1$ \Comment{full vertex set}
\For{$k = n_v$ \textbf{down to} $2$}
    \ForAll{$S \in \mathrm{Den}_k$}
        \State $G[S] \gets \dfrac{1}{\Sigma(\mathrm{Den}_k[S])}\displaystyle\sum_{j} h\big(\mathrm{Num}_k[S][j]\big)\; G\big[\mathrm{Succ}_k[S][j]\big]$
    \EndFor
\EndFor
\State $L \gets n-n_v+1$ \Comment{$L=n_v+1$ for a two-body diagram}
\If{fermi}
    \State $\mathcal{I} \gets \dfrac{G[\{r\}]}{2^{\,L}}\displaystyle\prod_{e=1}^{n}\dfrac{1}{\cosh\!\big(\beta\varepsilon_e/2\big)}$
\Else
    \State $\mathcal{I} \gets \dfrac{G[\{r\}]}{2^{\,L}}\displaystyle\prod_{e=1}^{n}\dfrac{1}{\sinh\!\big(\beta\varepsilon_e/2\big)}$
\EndIf
\State \Return $\mathcal{I}$
\end{algorithmic}
\end{algorithm}

Because the forward stage is independent of the energies, it is executed once per diagram, and only the backward stage must be repeated for each set of energies, as in a Monte Carlo sampling. The cost of both stages and the number of generated terms are discussed in Sec.~\ref{sec:complexity}. The algorithm assumes a connected diagram without self-loops. In the backward stage the denominators must be non-zero, so exactly degenerate energies require separate treatment. For bosonic lines, a line with zero energy is a genuine pole of the sum.

\section{\label{sec:complexity}Complexity}
Consider a connected two-body vacuum diagram with $n_v$ vertices and $n=2n_v$ edges, which has $L=n_v+1$ independent Matsubara frequencies. The algorithm does not enumerate the $n_v!$ time orderings of the vertices. Instead, it builds the result recursively over connected vertex subsets that contain a fixed root vertex. Each such subset (a state) is identified with an $n_v$-bit mask. Because the root is fixed, each of the remaining $n_v-1$ vertices is either inside or outside the subset, so the number of states is at most $2^{\,n_v-1}$. Excluding the full vertex set, which needs no denominator, the number of distinct denominators is at most $2^{\,n_v-1}-1$. Each state requires one denominator, namely a signed sum of the energies of the propagators crossing the cut.
\begin{figure}[ht]
        \centering
        \includegraphics[width=0.65\columnwidth]{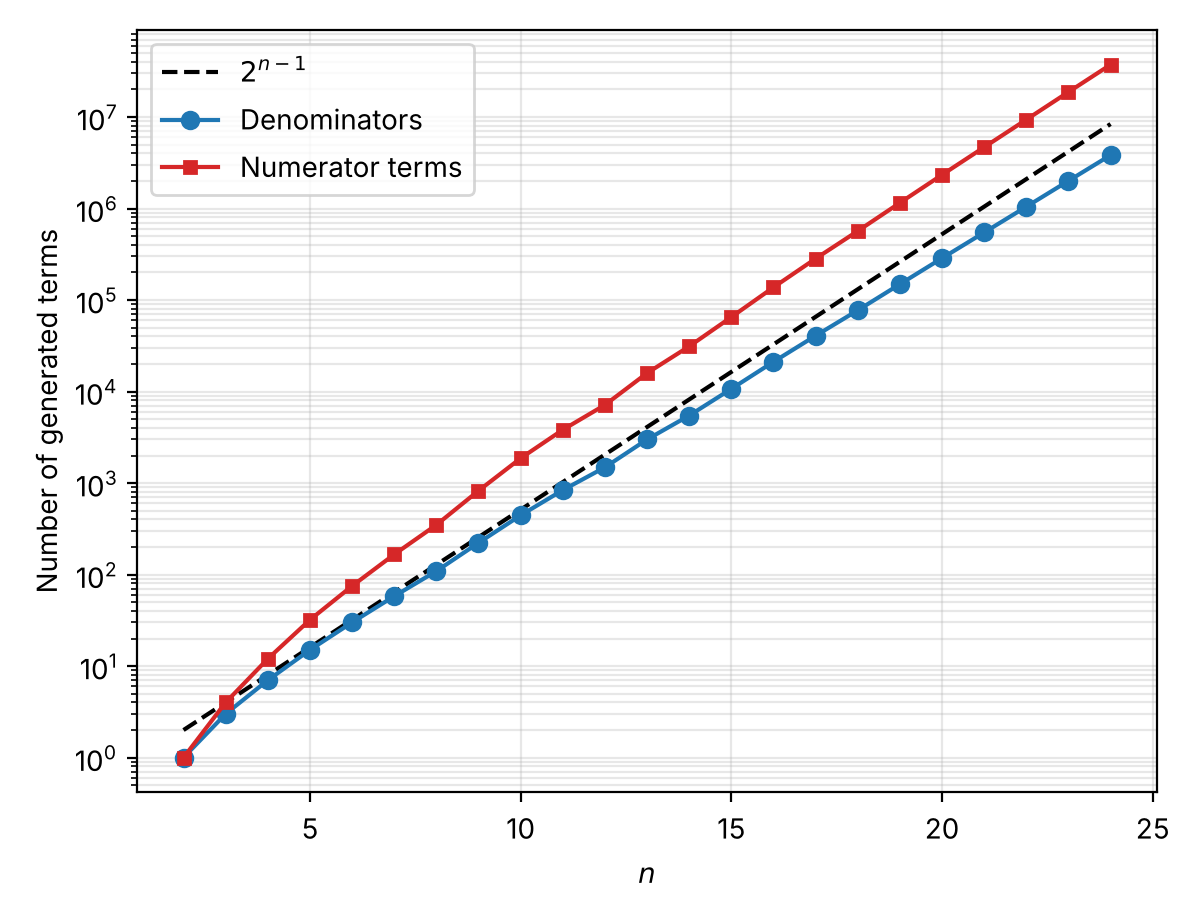}
        \caption{Number of denominators (blue) and numerator terms (red) generated by the subset recursion for the family of worst-case diagrams, $n_v=2$--$24$, compared with $2^{n_v-1}$ (dashed), on a logarithmic scale.}
        \label{generatedTerms}
\end{figure}

From each state, one transition is generated for every neighbouring vertex that can be added next, and each transition contributes one numerator, a hyperbolic function of the signed sum of the energies of the propagators joining the added vertex to the subset. A subset of $k$ vertices has at most $n_v-k$ such transitions, and summing over all subsets containing the root gives at most $(n_v-1)\,2^{\,n_v-2}<n_v\,2^{\,n_v-1}$ numerator terms. Both the cut energy of a subset and the energies attached to an added vertex can be updated incrementally from the four propagators of the added vertex, so the arithmetic cost is
\begin{equation}
    O\!\left(n\,2^{\,n_v-1}\right),
\end{equation}
where $n=2n_v$ is the number of edges, and the memory cost is $O(2^{\,n_v-1})$.

\begin{figure}[H]
        \centering
        \includegraphics[width=0.65\columnwidth]{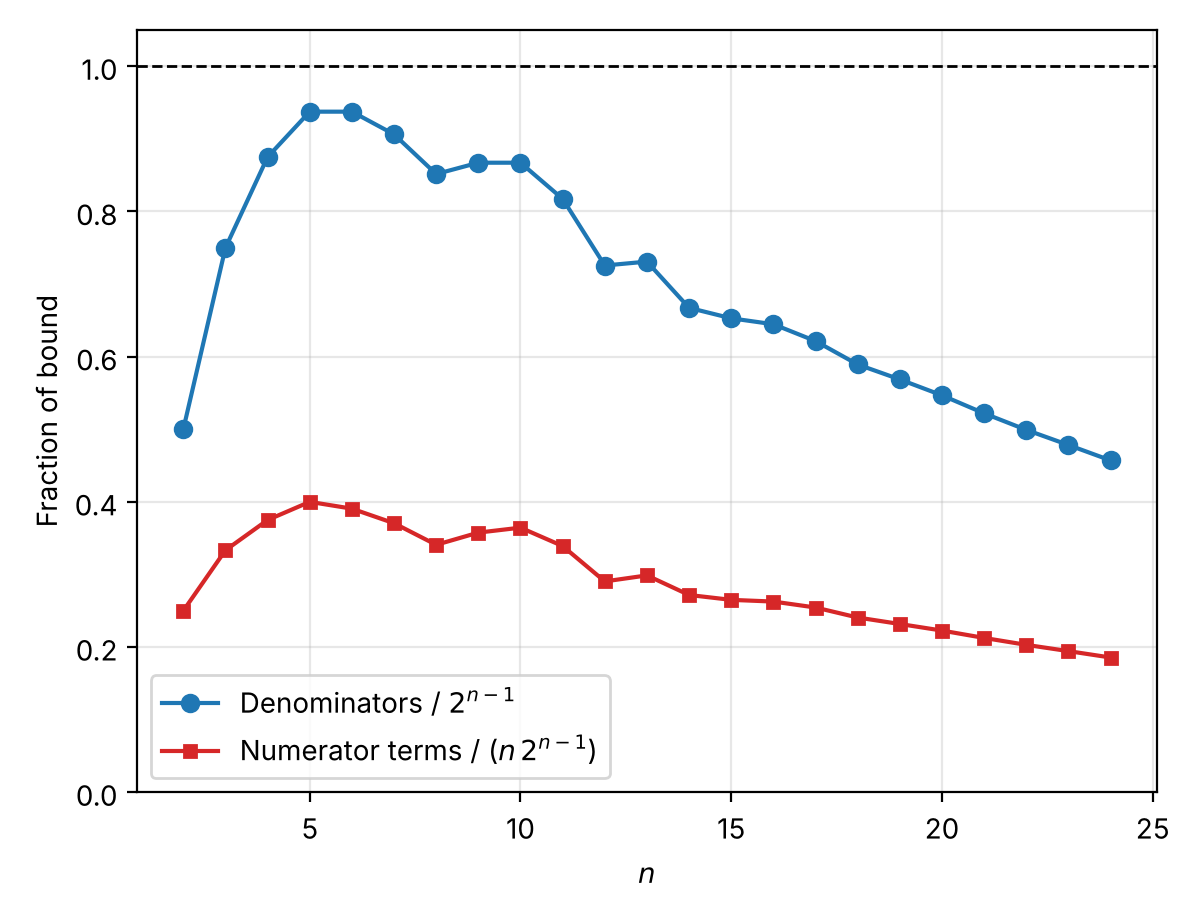}
        \caption{Number of denominators and numerator terms from Fig.~\ref{generatedTerms}, expressed as fractions of the bounds $2^{n_v-1}$ (denominators) and $n_v\,2^{n_v-1}$ (numerators). The number of denominators reaches its maximum possible value, $2^{n_v-1}-1$, for $n_v\le5$, and falls to $0.46\times2^{n_v-1}$ at $n_v=24$ ($3.8\times10^{6}$ denominators, $3.7\times10^{7}$ numerator terms).}
        \label{fractionOfBound}
\end{figure}

Figure~\ref{generatedTerms} shows the number of denominators and numerator terms actually generated for the family of worst-case diagrams, for $n_v=2$--$24$. On the logarithmic scale both curves are close to straight lines parallel to $2^{\,n_v-1}$, as expected for exponential growth with a base not larger than $2$, and the number of numerator terms exceeds the number of denominators by about one order of magnitude at $n_v=24$. The fractions of the bounds are shown in Fig.~\ref{fractionOfBound}. For $n_v\le5$ every subset containing the root is connected, and the number of denominators equals its maximum possible value $2^{\,n_v-1}-1$. For larger $n_v$ it remains below $2^{\,n_v-1}$ and the fraction decreases, from $0.94$ at $n_v=6$ to $0.46$ at $n_v=24$, where $3.8\times10^{6}$ denominators and $3.7\times10^{7}$ numerator terms are generated. The numerator terms follow the same trend relative to $n_v\,2^{\,n_v-1}$, falling from $0.40$ at $n_v=5$ to $0.19$ at $n_v=24$. The ratio of successive denominator counts at $n_v=24$ is about $1.9$. These observations are for the tested family only and do not constitute a bound on the growth rate. Finally, the construction of the states and transitions depends only on the topology of the diagram and not on the propagator energies, so it can be done once per diagram and reused for any number of energy sets.

\section{Conclusion}

We have introduced the Contraction Algorithm for Matsubara Sums (CAMS), a recursive vertex-contraction method that eliminates the internal Matsubara frequencies of connected Feynman or Hugenholtz diagrams. The result is a closed expression in the propagator energies and the inverse temperature, obtained from the recursion $G(S)=\Delta E_r(S)^{-1}\sum_{v}K(v_r,v)\,G(S\cup\{v\})$. The same rule covers fermionic lines, where the numerator is a hyperbolic cosine or sine depending on the parity of the number of connecting lines, and bosonic lines, where it is always a hyperbolic sine. External frequencies enter the denominators, so self-energy-type diagrams are treated in the same framework.

The recursion runs over connected vertex subsets that contain a fixed root vertex, and therefore avoids the enumeration of the $n_v!$ time orderings. For a two-body diagram with $n_v$ vertices and $n$ edges, the number of denominators is at most $2^{\,n_v-1}-1$, the arithmetic cost is $O(n\,2^{\,n_v-1})$, and the memory cost is $O(2^{\,n_v-1})$. Because the forward stage depends only on the topology, it is executed once per diagram, and only the backward evaluation is repeated for each set of energies, as in Monte Carlo sampling.

The method has limitations. Diagrams that mix fermionic and bosonic internal lines are not yet covered and require per-line statistics. Future work includes this extension, a systematic treatment of degenerate energies, and the use of CAMS in Monte Carlo evaluations of thermodynamic quantities.

\bibliography{references}

\end{document}